\documentclass[aps,twocolumn]{revtex4}

\begin{document}

\title{Prospects for making polar molecules with microwave fields.}

\author{Svetlana Kotochigova}

\address{Department of Physics, Temple University, Philadelphia, PA 19122-6082, USA}

\begin{abstract}

We propose a new mechanism to produce ultracold polar molecules with microwave fields. 
The proposed mechanism converts trapped ultracold atoms of different species into vibrationally excited molecules by a single microwave transition and entirely depends on the existence of a permanent dipole moment in the molecules. 
As opposed to production of molecules by photoassociation or magnetic-field  Feshbach resonances our method does not rely on the structure and lifetime of excited states or existence of Feshbach resonances. In addition, we determine conditions for optimal creation of polar molecules in vibrationally excited states of the ground-state potential by changing  frequency and intensity of the microwave field.  We also explore the possibility  to produce vibrationally cold molecules by combining the microwave field with an optical Raman transition or by applying a microwave field to Feshbach molecules. The production mechanism is illustrated for two polar molecules: KRb and RbCs.  

\end{abstract}

\maketitle

Quantum gasses of polar molecules have attracted attention of researchers due to their exceptional properties that allow them to interact over large distances by the dipole-dipole force. These properties  might be observable  when polar molecules are produced at ultra-cold temperatures.  To date the coldest polar molecules (with a temperature below 1 $\mu$K) have been created by either magnetic-field Feshbach  resonances  \cite{Donley,Regal,Strecker,Herbig}, photoassociation \cite{Lett,Julienne,Stwalley,Fioretti,Wynar,Kerman,Tiesinga} or, most recently, radio-frequency magnetic-dipole transitions \cite{Emst}. In the majority of cases these molecules are formed in highly-excited vibrational levels of the electronic ground state. However, many proposals to use polar molecules, either in simulation of many-body systems \cite{Baranov,Goral,Damski} or as qubits in a quantum processor \cite{DeMille}, require these molecules to be vibrationally cold.  The latter was first achieved in the photoassociation experiment by DeMille's group at Yale University \cite{Sage}. They formed polar RbCs molecules in the $v$ = 0 level of the X$^1\Sigma^+$ ground electronic potential, although the number of vibrationally cold molecules was relatively small and probably insufficient for the above mentioned applications. The small number of molecules might have been due to a significant loss of molecules from atom-molecule and molecule-molecule collisions in the trap as well as from uncontrolled spontaneous emission from the electronically-excited molecules. 

The latest developments in photoassociation, which show great promise to increase the molecular production rate,  is the use of an optical lattice. Experiments \cite{Grener} have demonstrated that a Mott insulator phase can be formed with exactly two atoms per optical lattice site. Atoms can then be associated into a single molecule in each lattice site.  The advantage of using an optical lattice is that a pair of atoms in a site is well isolated from distractive perturbations by the mean field of other atoms or molecules before or after the association process. In addition, a strong confinement of the atomic pair decreases the separation between the atoms so that Franck-Condon factors for association are expected to increase. The first photoassociation experiments in a lattice were done with homonuclear Rb$_2$ molecules \cite{Rom,Thalhammer}. Polar KRb molecules have been produced in an optical lattice by using magnetic-dipole transitions \cite{Emst}. 

\begin{figure}
\vspace*{5cm}
\includegraphics{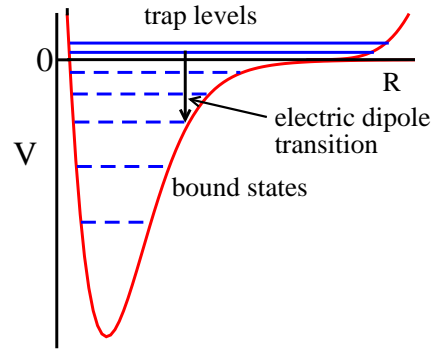}
\caption{Our association scheme to create polar molecules with microwave fields.  The schematic drawing shows the ground state potential plus an external harmonic trapping potential for two atoms of different species as
a function of interatomic separation $R$. The nonzero permanent dipole moment in polar molecules allows electric-dipole transitions between trap levels and vibrational levels (bound states) of the ground potential.}
\label{scheme} 
\end{figure}
 
In this Letter we propose a new approach to create polar molecules by using a microwave field.
Molecules can be produced by a single electric-dipole transition in excited vibrational states of the ground configuration 
as schematically shown in Fig.~\ref{scheme}.  For this proposal we will use the unique property of polar molecules, the existence of a nonzero permanent dipole moment,  which allows electric-dipole transitions from trap levels of optically confined atoms to excited ro-vibrational levels or between ro-vibrational levels of the ground electronic potentials. We believe the proposed method has advantages over other association methods. First of all, it explores transitions within the vibrational levels of the ground state potential, therefore  does not rely on the structure and lifetime of the excited states potential as it does for the photoassociation (PA) method. Secondly, it is not restricted to atomic systems that have magnetic Feshbach resonances. On the other hand, our proposal
can only be used with heteronuclear molecules, which posses the permanent dipole moments. There are, however, many such molecules.
Moreover, we might envision to binding molecules together by microwave-frequency electric-dipole transitions to create even bigger molecules or to
making production of molecules spatially dependent  by applying  additional spatially-dependent static electric or magnetic  fields. 

The bottleneck in any molecule forming process from its constituent atoms is the transition from two free atoms to a weakly-bound ``floppy'' molecule. As we will show later, the molecular production rates are such that our method as well as any other association method will benefit from confining the constituent atoms in an optical lattice. 
 
\begin{figure}
\vspace*{6.5cm}
\includegraphics{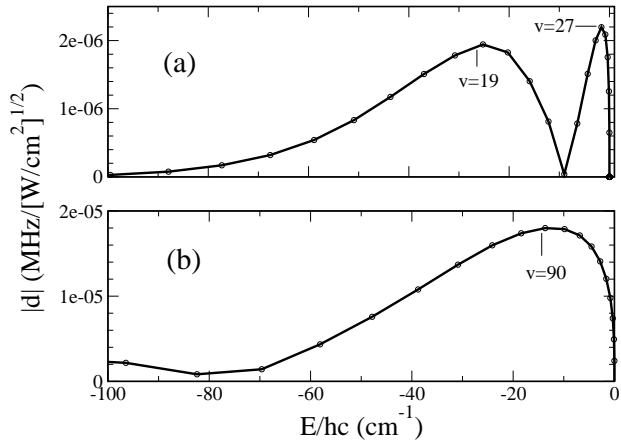}
\caption{The vibrationally-averaged dipole moment of the KRb molecule for transitions from the first trap level of a harmonic potential with a frequency of $\omega/2\pi$ = 200 kHz to vibrational levels of the a$^3\Sigma^+$ state (panel (a)) and X$^1\Sigma^+$ state (panel (b)) as a function of binding energy of the final state.}
\label{trap_KRb}
\end{figure}

In this Letter we focus on the detailed numerical verification of the proposed association scheme for KRb and RbCs. 
We assume that  two atoms of different species are confined in sites of an optical lattice and prepared in the lowest motional state of these sites. For simplicity we assume that this confinement can be described by a spherically symmetric harmonic trap with frequency $\omega$. This also leads to a harmonic trapping potential for the relative motion between atoms. Such a trapping potential must be added to and modifies the long-range behavior of the molecular potentials. 
For our calculation the frequency of the trapping potential is set to $\omega/2\pi$ = 200 kHz. The lowest harmonic oscillator levels are drawn schematically as the solid lines in Fig.~\ref{scheme}. 

We use a microwave field to induce a electric-dipole transition and to associate the lowest and rotationless trap level into a weakly-bound vibrational level with one unit of rotational angular momentum of the a$^3\Sigma^+$ or X$^1\Sigma^+$ state. The transition moment is determined by the matrix element 
\begin{equation}
d = C\int_0^\infty dR\, \phi_{v,\ell=1}(R) d(R) \phi_{\rm trap,\ell=0}(R)\, 
\end{equation}
where $R$ is the interatomic separation, $d(R)$ is the permanent electronic dipole moment of the  a$^3\Sigma^+$ or X$^1\Sigma^+$ states of  KRb and RbCs obtained from Ref.~\cite{Kotochigova,Kotochigova1}. The wavefunction  $\phi_{\rm trap,\ell=0}$ is the lowest harmonic oscillator state of either the a$^3\Sigma^+$ or X$^1\Sigma^+$ potential plus a harmonic potential. The function $\phi_{v,\ell=1}$ is a $\ell=1$ or $p$-wave rovibrational wavefunction of the same ground state potential as the initial state with binding energy $E_{v\ell}$. The dimensionless coefficient $C$ contains the angular dependence of the transition and in principle depends on the magnetic sublevels as well as the polarization of the microwave photon. It is always of the order of 1. For this
Letter we will assume it is equal to 1.
Furthermore, for the purposes of this study the hyperfine interaction and relativistic splitting between the $\Omega=0^-$ and 1 components of the a$^3\Sigma^+$ potential can be neglected. For an accurate prediction of binding energies of the ground state potentials  we needed to combine the best electronic potentials available \cite{Kotochigova2, Kotochigova1}, RKR data \cite{Amiot,Allouche,Fahs}, and long-range dispersion coefficients \cite{Porsev,Derevianko}. The theoretical electronic potentials have been slightly modified to fit to experimental measurements of scattering lengths where available \cite{Ferlaino,Tiesinga1}.  

\begin{figure}
\vspace*{6.5cm}
\includegraphics{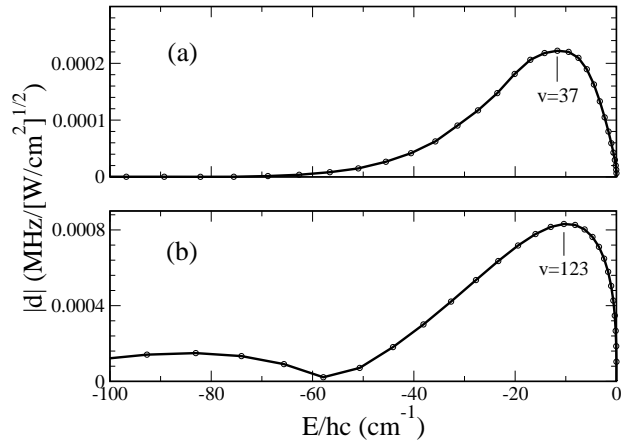}
\caption{The vibrationally-averaged dipole moment of the RbCs molecule for transitions from the first trap level of a harmonic potential with a frequency of $\omega/2\pi$ = 200 kHz to vibrational levels of the a$^3\Sigma^+$ state (panel (a)) and X$^1\Sigma^+$ state (panel (b)) as a function of binding energy of the final state.}
\label{trap_RbCs}
\end{figure}

Figures~\ref{trap_KRb} and \ref{trap_RbCs} show the absolute value of this vibrationally averaged transition dipole moment from the first trap level to the vibrational levels of the same potential for KRb and RbCs, respectively. In both figures the horizontal axis corresponds to the binding energies of the final state, which can also be interpreted as the microwave frequency needed to make the transition from the trap level. The markers on the curves in the figures correspond to bound vibrational states of the potentials. Uncertainties in the binding energies will not affect the functional shape of transition dipole moment. In Fig.~\ref{trap_KRb} the maximum association dipole moment  for the a$^3\Sigma^+$ potential occurs for vibrational levels $v$ = 27  and $v$ = 19 bound by about $-$1.5 cm$^{-1}$ and  $-$25 cm$^{-1}$, respectively,  and for vibrational level  $v$=90  bound by about $-$13 cm$^{-1}$ for the  X$^1\Sigma^+$ potential. In Fig.~\ref{trap_RbCs} the maximum association dipole moment occurs  for the $v$=37 and $v$=123 vibrational levels at a binding energy of approximately $-$10 cm$^{-1}$ for both the a$^3\Sigma^+$ and  X$^1\Sigma^+$ potentials, respectively.  

\begin{figure}
\vspace*{7cm}
\includegraphics{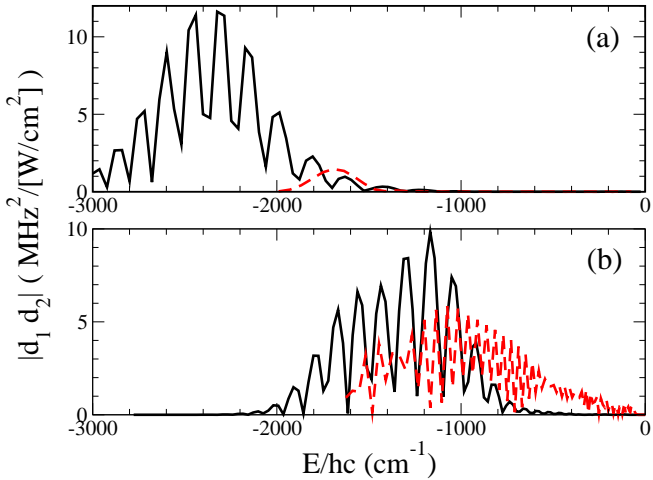}
\caption{The two-photon vibrationally-averaged dipole moments of KRb (panel(a)) and RbCs (panel (b)) for the optical Raman transition as a function of vibrational binding energies of the 3(1) (solid line) and 4(1) (dashed line) intermediate states . The panel (a) corresponds to the pathway from $v$=27  of  the 
a$^3\Sigma^+$ potential to the $v$=0, $\ell$=1 X$^1\Sigma^+$ ground state of KRb. The panel (b) corresponds to the pathway from $v$=37  of  the a$^3\Sigma^+$ potential to the $v$=0, $\ell$=1 X$^1\Sigma^+$ ground state of  RbCs.}
\label{raman}
\end{figure}

The vibrationally averaged dipole moments are not the only relevant quantities for determining a successful association mechanism. Once a molecule has been formed the same microwave field can transfer it into trap levels other than the lowest. This could limit a number of molecules that can be made. However, in strongly confined traps this effect can be made small.  For  field intensities below $I = 10$ kW/cm$^2$ the Rabi frequency, which is in units of MHz given by $\Omega$ = $d ({\rm MHz}/\sqrt{{\rm W/cm}^2}) \times \sqrt{I({\rm W/cm}^2)}$,  of the microwave transition is smaller than 100 kHz and thus less than our harmonic trapping frequency of $\omega/2\pi$ = 200 kHz. As a result we will have a perfect two level system. This will prevent loss of molecules due to population of higher trap levels during Rabi flopping.

We also find  that the transition dipole is proportional to $\omega^{3/4}$, or alternatively as inversely proportional to the square root of the volume of the trap state. A tighter trap will increase the transition dipole moment. The proportionality is consistent with the Wigner threshold behavior for vibrationally averaged transition dipole moments. In other words, association of molecules in optical lattices is favorable. 

References \cite{Kotochigova3,Kotochigova1} have shown that the room-temperature black-body and natural life-time of vibrational levels of the a$^3\Sigma^+$ and  X$^1\Sigma^+$ states of KRb and RbCs are at least as large as $10^6$ s.   This large life-time of vibrational levels of the a$^3\Sigma^+$ and  X$^1\Sigma^+$ potentials allows us sufficient time for the creation of a dense cloud of molecules in an optical lattice. 

The dipole moments in Figs. \ref{trap_KRb} and \ref{trap_RbCs} are rather small compared to the dipole moments of first step of optical photoassociation. The latter  correspond to transitions from a trap level to high vibrational levels of electronically excited potentials \cite{Jaksch}. However, this is only half of this molecular formation process. Often the second step consists of a spontaneous decay into many rotational and vibrational levels of the ground state potential. For example, experiments \cite{Kerman,Sage} have produced $v$=37 a$^3\Sigma^+$ molecules of RbCs by spontaneous decay, using a transition via an intermediate excited potential, which dissociates to the Rb $^2P_{1/2}$ atomic limit. They determined that only 7\% of the excited molecules decay into the $v$=37 level. Our estimate of this process shows that the total transition dipole moment has the same order of magnitude as our microwave process.
 
It is often preferable for  molecules to be vibrationally cold. We propose that  the microwave transition will be followed by one optical Raman transition to form the vibrationally cold molecules. Hence, we apply a microwave field to accumulate a sufficient number of molecules in one excited vibrational level of either the a or X potential, and then use an optical Raman transition to convert these excited molecules into vibrationally cold $v=0, \ell=1$ X$^1\Sigma^+$ molecules. To show the strength of this scheme we calculate the effective dipole moment for the two-photon optical Raman transition, focusing on the  $v=27$ $\ell=1$  and $v=37$ $\ell=1$ vibrational levels of the a$^3\Sigma^+$ potential of KRb and RbCs, respectively, as starting levels of this process. These vibrational levels are chosen as they have the largest Rabi matrix elements with the lowest trap level. The $v$=37 level of RbCs was used in experiment \cite{Sage} to produce  the $v$=0  X$^1\Sigma^+$ molecules. 
The effective dipole moment for the two-photon Raman transition is $d_{\rm eff} = C'd_1 d_2/(\Delta + i\gamma/2)$, where $d_1$ and $d_2$ are the vibrationally-averaged transition dipole moments for the upwards and downwards transition, respectively. In this equation the dipole moments have units of MHz/$\sqrt{\rm W/cm^2}$. The quantities $\Delta$ and $\gamma$ in MHz are the detuning  from and the linewidth of a rovibrational level of the intermediate state. Finally, the dimensionless coefficient $C'$ contains all angular momentum information and is assumed to be 1.

The results of our calculation are shown in Fig.~\ref{raman} (a) and (b). The vertical axis gives the product of the absolute value of the upward and downward transition dipole and the horizontal axis shows the binding energies  of intermediate states relative to their own dissociation limit. In both cases we have used the vibrational levels of the 3(1) and 4(1) potentials as intermediate states as these are good candidates for molecular formation \cite{Kerman,Sage}.  A comparison of the transition rates determined from Fig.~\ref{raman}  with the more conventional photoassociation scheme of  the two-photon Raman transition from the trap level shows more than a three order of magnitude increase.

\begin{figure} 
\vspace*{4cm} 
\includegraphics{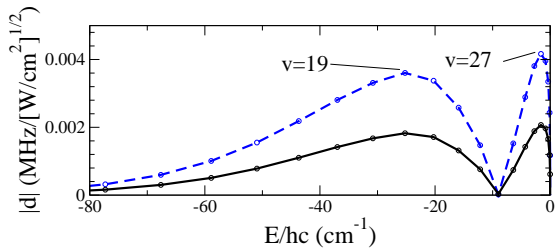} 
\caption{The vibrationally-averaged dipole moment for the transition from the $\ell$=0, $v$ = 31(solid line) and $\ell$=0, $v$ = 30 (dashed line) levels of the a$^3\Sigma^+$ of KRb to more deeply bound $\ell$ =1 vibrational levels of this state as a function of binding energy of the final state.} 
\label{rf} 
\end{figure}

We also propose to apply a single microwave electric-dipole transition to Feshbach molecules in order to convert these molecules into more deeply-bound states.  Feshbach molecules were successfully created in  KRb\cite{Inouye,Ferlaino} while so far non have been observed in RbCs. These molecules are produced by bringing a molecular level into resonance with an energy level of a pair of trapped atoms. This is achieved by slowly changing an external magnetic field. The molecular level is bound by no more than 0.1 cm$^{-1}$, a typical atomic hyperfine energy. In fact, this bound state is often the last or second from the last bound state of the a or X state.   Figure~\ref{rf} shows results of a calculation of  the transition dipole moment from the last and second from the last vibrational levels of the a$^3\Sigma^+$ potential of KRb to other vibrational levels. A comparison of Figs.~\ref{trap_KRb} and \ref{rf} shows a significant increase in the dipole moment.

In this Letter we have searched for routes to simple and controllable production of molecules in tightly-confining optical traps and optical lattices. As we have shown a microwave-driven molecular association method, which is unique to polar molecules, is one such option. This method does not suffer from uncontrollable spontaneous decay from the excited states. We also have shown
that molecules created by microwave transitions are easily converted to the vibrationally cold molecules.

The author acknowledges support from the Army Research Office.

\end{document}